\documentclass[conference]{IEEEtran}
\IEEEoverridecommandlockouts
\usepackage{cite}
\usepackage[hyphens]{url} 
\usepackage{amsmath,amssymb,amsfonts}
\usepackage{graphicx}
\usepackage{textcomp}
\usepackage{xcolor}
\usepackage{algorithm}
\usepackage{algorithmicx}
\usepackage{algpseudocode}
\usepackage{graphicx}
\usepackage{booktabs}
\usepackage{subcaption}
\usepackage{caption}
\usepackage{multirow}

\def\BibTeX{{\rm B\kern-.05em{\sc i\kern-.025em b}\kern-.08em
    T\kern-.1667em\lower.7ex\hbox{E}\kern-.125emX}}
\newcommand{\oursystem}{\textit{Iter-Q\textsuperscript{2}O}}

\begin{document}

\title{Towards a Hybrid Quantum-Classical Computing Framework for Database Optimization Problems in Real Time Setup}

\author{\IEEEauthorblockN{Hanwen Liu\textsuperscript{\textsection}}
\IEEEauthorblockA{\textit{Department of Computer Science} \\
\textit{University of Southern California}\\
Los Angeles, CA, USA \\
hanwen\_liu@usc.edu}
\and
\IEEEauthorblockN{Ibrahim Sabek\textsuperscript{\textsection}}
\IEEEauthorblockA{\textit{Department of Computer Science} \\
\textit{University of Southern California}\\
Los Angeles, CA, USA \\
sabek@usc.edu}
} 

\maketitle

\begingroup\renewcommand\thefootnote{\textsection}
\footnotetext{Both authors have equal contributions, and their names are sorted alphabetically.}
\endgroup

\begin{abstract}
Quantum computing has shown promise for solving complex optimization problems in databases, such as join ordering and index selection. Prior work often submits formulated problems directly to black-box quantum or quantum-inspired solvers with the expectation of directly obtaining a good final solution. Due to the black-box nature of these solvers, users cannot perform fine-grained control over the solving procedure to balance the accuracy and efficiency, which in turn limits flexibility in real-time settings where most database problems arise. Moreover, it leads to limited potential for handling large-scale database optimization problems. In this paper, we propose a vision for the first real-time quantum-augmented database system, enabling transparent solutions for database optimization problems. We develop two complementary scalability strategies to address large-scale challenges, overcomplexity, and oversizing that exceed hardware limits. We integrate our approach with a database query optimizer as a preliminary prototype, evaluating on real-world workload, achieving up to 14x improvement over the classical query optimizer. We also achieve both better efficiency and solution quality than a black-box quantum solver.
\end{abstract}

\section{Introduction}
\label{sec:introduction}

Database systems encompass multiple optimization problems, such as join ordering (e.g.,~\cite{qosurvey96,joref25,bao,yu2020reinforcementlstm,rejoin}), transaction scheduling (e.g.,~\cite{transsched24,learntranssched25}), multi-query optimization (e.g.,~\cite{mqo88,mqoref17}), and index tuning~\cite{indextune1,indextune2,learnindextune22}. The quality of their solutions is critical to overall performance. As data volumes grow exponentially and query workloads become more complex, these optimization tasks  have become increasingly difficult to solve efficiently~\cite{dbtune02}. These problems are inherently combinatorial, and for decades, database systems have relied on heuristic and other classical optimization techniques (e.g.,~\cite{qosurvey96,joref25,transsched24,mqo88,mqoref17,indextune1,indextune2}) that balance solution quality and computational efficiency. While such methods are effective in practice, they become less reliable when the search space is large, highly-complex, or must be explored under real-time or adaptive conditions. In these scenarios, the objective often shifts from finding the optimal solution to producing a high-quality solution quickly, where accepting a suboptimal result is justified if it substantially reduces optimization time.

A prime example is the join order optimization problem, a key challenge in query optimization that is NP-hard~\cite{cluet1995complexity} due to the exponential growth of the search space with the number of relations. Dynamic programming (DP)-based approaches (e.g.,~\cite{accesspath}) can compute optimal results by exhaustively enumerating all possible join sequences for small queries (i.e., queries with relatively few relations) but become impractical for larger ones. To address this limitation, a variety of heuristic methods (e.g.,~\cite{10.1145/1270.1498, accesspath, moerkotte2008dynamic, neumann2018adaptive}) have been proposed to balance solution quality and computational efficiency.

More recently, machine learning based techniques have emerged as promising alternatives for database optimization (e.g.,~\cite{bao,yu2020reinforcementlstm,rejoin,learntranssched25,learnindextune22,lsched,liu2025serag,liu2025sefrqo}). However, they continue to face significant challenges in hard combinatorial settings, including cold-start issues (e.g.,~\cite{bao}), dependence on large training datasets (e.g.,~\cite{neo}), and the need for frequent retraining as workloads evolve (e.g.,~\cite{lsched}).

Fortunately, recent advances in quantum hardware (e.g.,~\cite{dwave5000,ibmquantum2022,rigetti2022}) have positioned quantum computing as a powerful and promising computational paradigm.  Grounded in widely accepted complexity theoretic assumptions~\cite{qcassume09,qcassume2,grovesearch96}, quantum systems are believed to surpass classical systems in computational capability, with the potential to (i)~achieve quantum advantage~\cite{googlebreakthrough,dwavebreakthrough,quantumadvantage} on specific problems by solving them faster than the best known classical algorithms, for example in cryptanalysis~\cite{crypto1,Zhao_2023} and simulation~\cite{googlebreakthrough,dwavebreakthrough}, and (ii) efficiently explore exponentially large search spaces, particularly for certain optimization problems~\cite{10821238,10247202}. Moreover, quantum computing is becoming increasingly accessible through cloud platforms such as D-Wave~\cite{dwaveleap} and IBM Quantum~\cite{ibmquantum2022}. Forward looking projections~\cite{googlequantum24,microsoftquantum25} anticipate rapid growth in quantum capacity and the emergence of on-premise hardware in the near future, enabling lower latency execution and tighter system integration.

Despite their transformative potential, quantum computing systems remain in an early stage of development and face significant hardware-related challenges. Current devices operate in the noisy intermediate-scale quantum (NISQ) regime and exhibit limitations such as restricted qubit counts, sparse connectivity, substantial noise, and the absence of robust error correction~\cite{preskill2018NISQ}. Quantum annealers~\cite{dwave5000}, such as those developed by D-Wave~\cite{dwaveleap}, have shown promise for combinatorial optimization in real-world applications (e.g., production scheduling~\cite{10821238}, power network optimization~\cite{10247202}) by exploiting quantum tunneling~\cite{razavy2013quantum} and parallel exploration of solution spaces. However, their limited qubit capacity and sparse hardware topologies (e.g., Pegasus~\cite{Pegasus}, Zephyr~\cite{zephyr}) constrain the size and density of problems that can be efficiently embedded. As a result, pure quantum annealing remains insufficient for solving complex, large-scale database optimization tasks (e.g.,~\cite{qdm26,qtrans20,qdb13,TK16,schonberger2023workshop,readytoleap,liu2026quantumcomputingreadyrealtime}).

To overcome these bottlenecks, hybrid quantum–classical computing has emerged as a practical paradigm in which quantum processing units (QPUs) accelerate only the most computationally intensive subproblems within broader classical optimization pipelines~\cite{hybridqcref21,dwavehss}. Such hybridization leverages the complementary strengths of both worlds: classical processors handle large-scale orchestration and problem decomposition, while quantum annealers focus on sampling and exploring complex energy landscapes that are otherwise computationally prohibitive. Yet, applying existing hybrid quantum–classical methods to database optimization introduces several unique challenges, especially in a \textit{real-time} setup:

\noindent\textbf{C1. Solver Opacity: Rely on black-box solvers.}
Most current research (e.g.~\cite{q2odemo25,hanwenabstract,sabekbushyjoin25}) relies heavily on commercial, black-box quantum solvers (e.g., D-Wave's CQM Solver~\cite{dwavehss} and NL-Solver~\cite{nlsolver25}), and on quantum-inspired solvers (e.g., Digital Annealer~\cite{aramon2019physics}). Without the required transparency from quantum solvers, researchers typically formulate optimization problems in the Quadratic Unconstrained Binary Optimization (QUBO) format and submit them directly to these solvers (e.g.,~\cite{DA,bushjointree}). This black-box usage restricts fine-grained user control over the quantum annealing process and limits flexibility for debugging and tuning. Consequently, it hinders the design of more efficient algorithms for database optimization in real-time settings.

\noindent\textbf{C2. QUBO Overcomplexity.}
Most existing approaches faithfully map the original problems into QUBO form~\cite{sabekbushyjoin25,transactionqubo,qtranssched25,TK16,fankhauser2021multiple,DA}. Although this preserves semantics, it produces overly complex QUBOs that yield dense interaction graphs and excessive auxiliary variables. This complicates variable-to-qubit mapping and slows down embedding on sparsely connected hardware, such as D-Wave's Pegasus~\cite{Pegasus} and Zephyr~\cite{zephyr}. Moreover, the added complexity degrades solver performance in finding high-quality solutions. Many database problems, however, exhibit redundancy or sparsity. For example, in join order optimization, if joins $A \bowtie B$ and $B \bowtie C$ are selected, there is no need to introduce additional variables to encode that tables $A$ and $C$ are connected~\cite{joinsearchspace01,sabekbushyjoin25}. Blindly preserving all correlations increases complexity without commensurate gains.

\noindent\textbf{C3. QUBO Oversize: Beyond hardware limits and semantics-aware decomposition.}
As database optimization problems grow, mapping a QUBO to physical qubits can exceed hardware capacity. Feasibility requires that every decision variable be mapped to a physical qubit. Problems beyond this threshold must be decomposed and later recomposed. Generic decomposition that ignores database semantics treats each QUBO as a standalone, independent problem. Missing structural semantic information forfeits potential simplifications during recomposition and limits adaptability when parameters change slightly. Practical scaling, therefore, demands decomposition that preserves task semantics and minimizes shared variables across subproblems, so that recomposition remains efficient and the final solutions remain valid.

To address these challenges, we propose a scalable hybrid quantum-classical framework for solving large-scale database optimization problems \textbf{in real-time settings}. Specifically, this framework \textbf{simplifies} (i.e., iterative correlation relaxation) or \textbf{partitions} (i.e., problem-aware decomposition) QUBO structures while preserving solution correctness. This paradigm directly leverages the lowest-level operations (i.e., sampling) on the quantum annealer, enabling transparent and fine-grained control to achieve a balance between solution quality and runtime efficiency. Our preliminary prototype and evaluation demonstrate up to a 14x improvement over the classical database component (i.e., PostgreSQL's query optimizer) and a commercial black-box quantum solver.

This paper presents three key takeaways:

\begin{enumerate}
    \item \textbf{New Perspective:} We introduce a vision for the \textit{first real-time quantum-augmented database system} that tightly integrates a hybrid quantum and classical optimizer, enabling transparent solutions for large-scale database optimization problems.
    \item \textbf{New Methods:} We develop two complementary scalability strategies. First, an iterative correlation relaxation framework that preserves the full set of variables while pruning and selectively reintroducing correlations based on semantics and their impact on solution quality. Second, a problem-aware decomposition–embedding–sampling–composition pipeline that preserves structure while partitioning QUBOs into semantically meaningful subproblems to solve large-scale optimization problems.
    \item \textbf{New Research Direction and Challenges:} We outline a promising hybrid quantum-classical framework for large-scale optimization that is deeply integrated within database systems, while noting that efficiency, simplicity, and generalizability remain ongoing challenges.
\end{enumerate}

This paper is organized as follows. Section~\ref{sec:preliminaries} introduces preliminaries on quantum annealing and essential background on current quantum computing methodology. Section~\ref{sec:system-overview} presents the system architecture and formalize the problem setting. Section~\ref{sec:scalable_hybrid_quantum_classical} details the two proposed methods. Section~\ref{sec:prototype} describes a preliminary prototype integrated with a database query optimizer. Section~\ref{sec:results} reports preliminary experimental results. Finally, Section~\ref{sec:conclusion} summarizes the findings and contributions of this vision paper.

\section{Background \& Preliminaries}\label{sec:preliminaries}

\begin{figure}[t]
    \centering
    \includegraphics[width=0.9\linewidth]{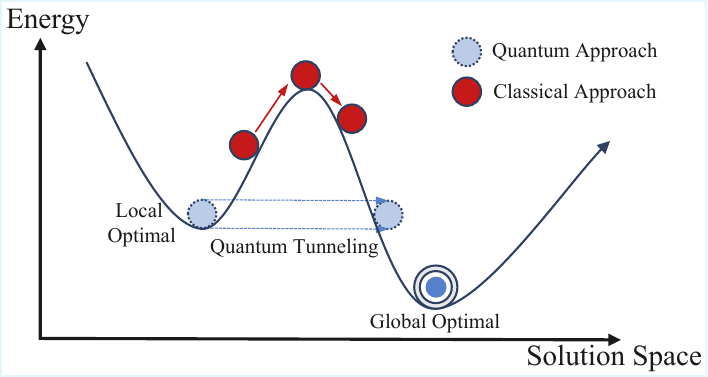}
    \caption{Quantum Annealing Compare with Classical Approach.}
    \label{fig:qa-compare}
\end{figure}

Quantum annealing (Section~\ref{subsec:quantum-annealing}) serves as the theoretical foundation and core technology in this work. It underpins the hybrid quantum classical optimization workflow (Section~\ref{subsec:hybrid-solver}). We then survey correlation relaxation for QUBO (Section~\ref{subsec:corr-relax}) and problem-aware hybrid optimization pipelines (Section~\ref{subsec:problem-aware}).

\subsection{Quantum Annealing}\label{subsec:quantum-annealing}
Quantum Annealing (QA) is a computational paradigm utilizing quantum hardware to solve complex optimization problems~\cite{kadowaki1998}. It shares similarities with classical simulated annealing (SA)~\cite{van1987simulated}: Both methods explore the search space by proposing random perturbations to the current state. Fig.~\ref{fig:qa-compare} compares SA and QA. In SA, a proposed move is accepted if it lowers the objective; otherwise, it is accepted with a probability that depends on the current temperature. Exploration in SA is driven by thermal fluctuations under a temperature schedule. 
In QA, the system evolves under a time-dependent Hamiltonian; the transverse field induces quantum fluctuations that promote exploration. Quantum tunneling further enables the system to traverse energy barriers rather than surmount them thermally, allowing escape from suboptimal states (i.e., avoid becoming trapped in local minima), thereby increasing the likelihood of converging to a global optimum.

QA solves problems that can be formulated as Quadratic Unconstrained Binary Optimization (QUBO)~\cite{Lucas2014}. In the QUBO form, QA seeks a binary assignment \(s\in\{0,1\}^n\) that minimizes the quadratic energy function~\cite{cqmjo24}:
\[
E(\vec{s}) \;=\; \sum_{i j} J_{i j}\, s_i\, s_j \;+\; \sum_{i} h_i\, s_i,
\]
where $s_i$ and $s_j$ are binary variables representing the qubit states, $h_i$ denotes the local bias on qubit $i$, and $J_{i j}$ represents the interaction strength between qubits. During quantum annealing, the initial superposition evolves toward a state that minimizes the system's energy, encoding the solution to the original combinatorial optimization problem.

\subsection{Hybrid Quantum-Classical Optimization}\label{subsec:hybrid-solver}
\begin{figure}[t]
  \centering
    \includegraphics[width=\linewidth]{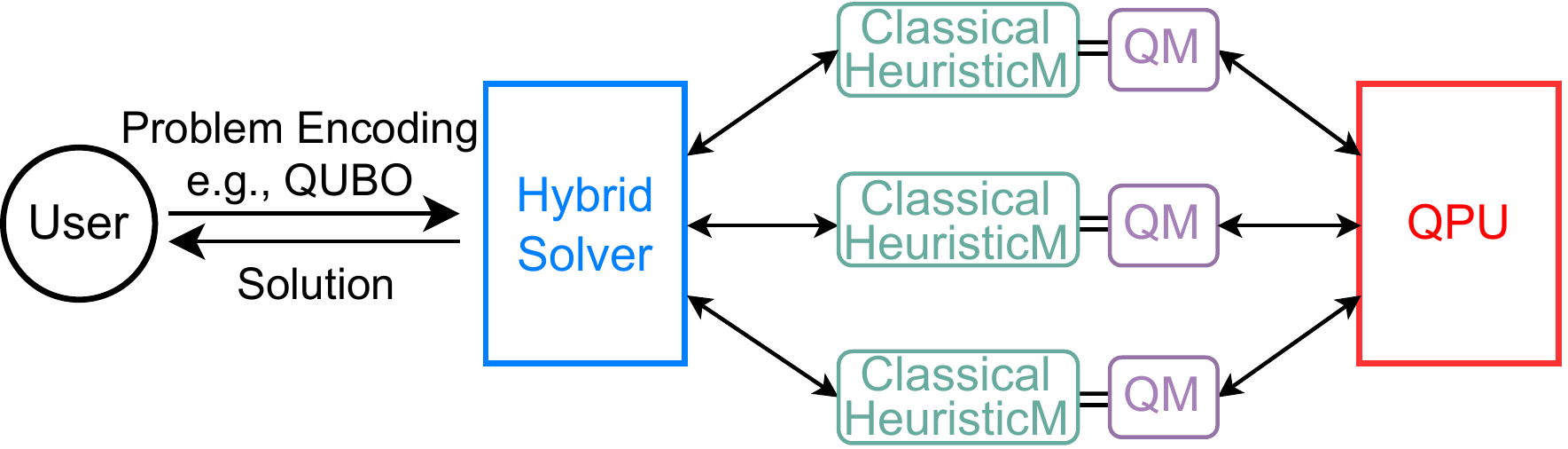}  
    \caption{A Generic Workflow for Hybrid Quantum-Classical Optimization.}
     \label{fig:dwave-hybrid-framework}  
\end{figure}

Hybrid quantum-classical optimization couples classical heuristics with quantum annealing (QA)~\cite{kadowaki1998} to improve scalability and solution quality. As shown in Fig.~\ref{fig:dwave-hybrid-framework}, the user encodes the optimization problem and submits it to a hybrid solver that launches parallel workers, each combining a classical heuristic module with a quantum module (QM). The heuristic module preprocesses, decomposes, and refines the search space, while the QM submits corresponding quantum queries to the Quantum Process Unit (QPU) and returns samples with energies; this feedback guides subsequent heuristic updates. The loop continues until a time budget or stopping criterion is met, after which the best solution is returned. Most commercial, black-box quantum solvers, such as D-Wave's CQM-Solver~\cite{dwavehss} and NL-Solver~\cite{nlsolver25}, follow this hybrid workflow.

\subsection{Correlation Relaxation for QUBO}\label{subsec:corr-relax}
Correlation relaxation simplifies the original QUBO while retaining the full set of decision variables. This technique has been explored in several hybrid optimization methods~\cite{qboost17,dwaveleap,dimod23}. A common strategy is to prioritize correlations with the largest energy contribution; in a QUBO representation, this amounts to ranking the pairwise terms in $J$ by their impact on the Hamiltonian energy~\cite{qdiscrete14}. However, a purely energy-based criterion does not carefully account for semantic correctness in database optimization problems, where correlations with lower energy may enforce critical semantics. For example, in join order optimization~\cite{DA,readytoleap}, a low energy correlation may encode a validity constraint that prevents duplicate joins or infeasible join structures, and pruning this correlation can yield an invalid plan. Similarly, in index tuning~\cite{qindextune24}, some low-weight constraints ensure that storage budgets are respected, and removing them can produce infeasible solutions. Therefore, energy-only correlation pruning, which does not consider semantic meaning, is insufficient for database optimization.

\subsection{Problem-aware Hybrid Optimization Pipelines}\label{subsec:problem-aware}
Commercial quantum providers (e.g., D-Wave~\cite{dwaveleap}, IBM~\cite{qiskit23}, Fujitsu~\cite{fujitsu2023}) offer hybrid optimization pipelines. Although they permit limited user customization, these pipelines rely on generic techniques, such as minor embedding~\cite{dwaveadvantage} and unconstrained merging~\cite{dwaveleap}, which do not account for problem-specific semantics, thereby limiting their effectiveness for database optimization use cases. To be effective in practice, decomposition must leverage problem-specific characteristics rather than generic QUBO graph properties, minimizing shared variables across subproblems to avoid recomposition bottlenecks. Few prior quantum-classical database optimization efforts \cite{qtranssched25,bushjointree,qmqo26} have incorporated problem characteristics only at the logical level (e.g., pruning unlikely join trees~\cite{bushjointree} or low-contention transactions~\cite{qtranssched25}) prior to QUBO construction.
While this reduces size, it treats the QUBO structure itself as a black box, failing to account for how database semantics map to embeddings and solver configurations. As a result, these solutions are not reusable or adaptable to workload shifts. For example, aggressive pruning of join trees might exclude query plans that could later become optimal under changing workloads.

\section{System Overview}\label{sec:system-overview}
In this section, we first present an overview of our vision for a quantum-augmented database system and its key components (Section~\ref{subsec:system-framework}). We then formally define the problem setting for achieving the best solution accuracy under a given time constraint, which guides the application of our system (Section~\ref{subsec:problem-formulation}).

\subsection{System Architecture}\label{subsec:system-framework}
\begin{figure}[t]
    \centering
     \includegraphics[width=1\linewidth]{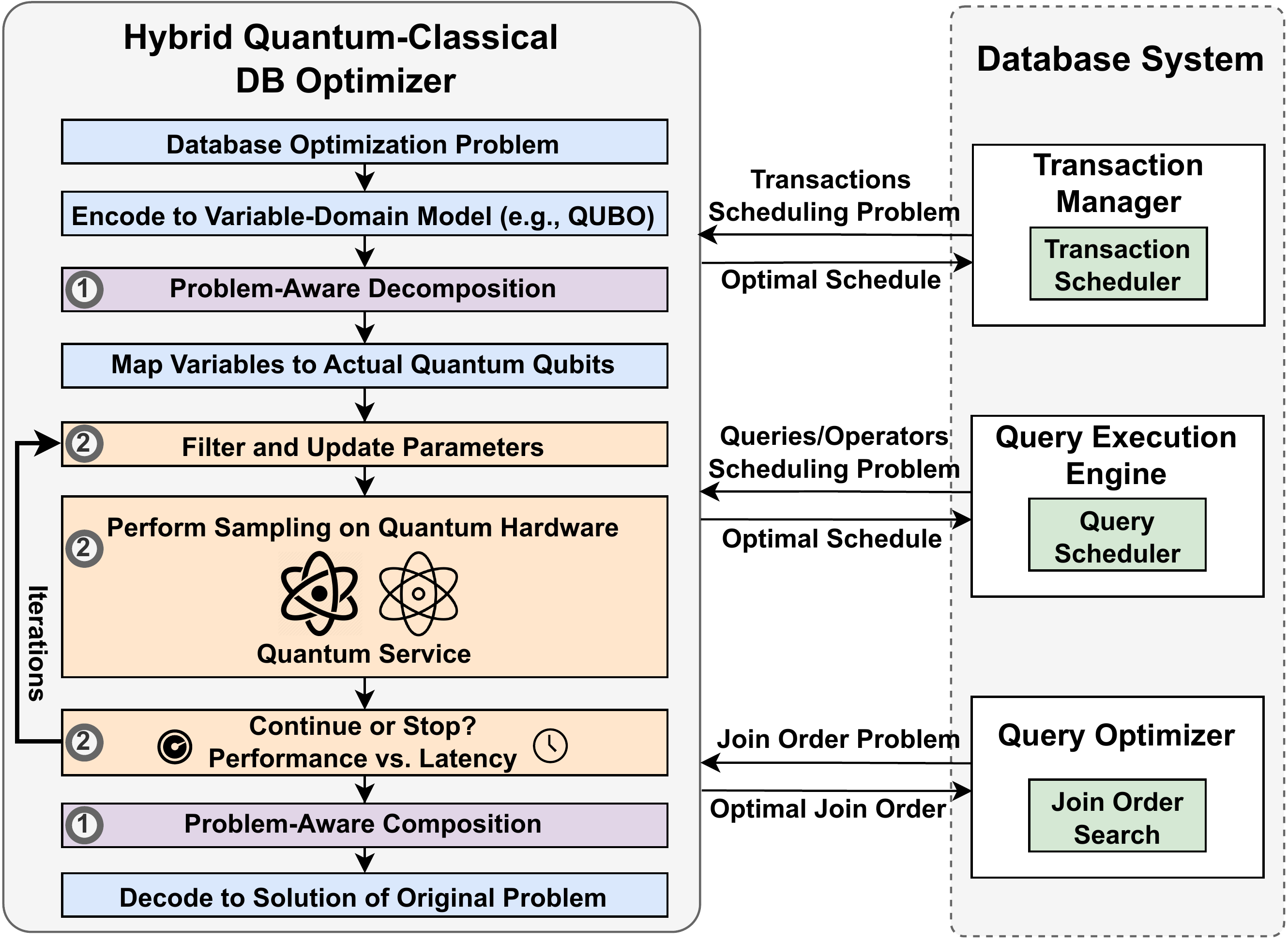}
    \caption{Overview of a Quantum-Augmented Database System.}
    \label{fig:system_overview}
\end{figure}

Figure~\ref{fig:system_overview} presents an overview of our vision for quantum-augmented database systems. We design a hybrid quantum-classical database optimizer that interfaces closely with core DBMS components, including the transaction manager, the query optimizer, and the execution engine. This visionary system resolves the three challenges in solving the large-scale database optimization problem (Check Section~\ref{sec:introduction}). 

In the workflow of the hybrid quantum-classical DB optimizer, the system first receives a large-scale database optimization problem, such as transaction scheduling, query or operator scheduling, or the join order problem. The problem is then encoded into a variable domain model, e.g., a QUBO. If the QUBO contains more variables than the capacity of the quantum annealer (Challenge C3), the system applies problem-aware decomposition (details in Section~\ref{subsec:research_task_2}). Afterward, the variables are mapped to physical quantum qubits as the initial step for execution on quantum hardware. If the QUBO is overly dense (Challenge C2), correlation relaxation (details in Section~\ref{subsec:research_task_1}) is applied. The quantum sampling, which is the lowest-level and fundamental operation, is then executed on the quantum hardware. Based on solution quality and user-defined timing constraints, our system determines whether to invoke the iterative correlation relaxation mechanism (details in Section~\ref{subsec:research_task_1}) to further improve solution quality. If so, the obtained samples and their energies are used to update the QUBO parameters, and the process continues. Otherwise, the reverse of decomposition, problem-aware composition (details in Section~\ref{subsec:research_task_2}) is performed, to produce the final QUBO solution. Finally, it is then decoded to obtain the solution to the original problem. This end-to-end workflow does not rely on any black-box quantum solver (Challenge C1) and provides transparent control over quantum sampling for solving large-scale optimization problems on quantum hardware.

\subsection{Problem Statement}\label{subsec:problem-formulation}
\newcommand{\DB}{\mathcal{D}}
\newcommand{\q}{\mathcal{Q}}
\newcommand{\Schema}{\mathcal{S}}
\newcommand{\Instance}{\mathcal{I}}
\newcommand{\Engine}{\mathcal{E}}
\newcommand{\Quality}{\mathrm{Qual}}
\newcommand{\Performance}{\mathrm{Perf}}
\newcommand{\QPUTime}{\mathrm{Time}_{\mathrm{QPU}}}
\newcommand{\QunatumTime}{\mathrm{Time}_{\mathrm{Quantum}}}
\newcommand{\SyntaxValid}{\mathrm{Syn}}
\newcommand{\SemEq}{\mathrm{SemEq}}
\newcommand{\Latency}{\mathrm{Latency}}
\newcommand{\Cost}{\mathrm{Cost}}
\newcommand{\semval}[2]{\llbracket #1 \rrbracket_{#2}}

We consider optimization tasks for executing a query $\mathcal{Q}$ over a database $\DB$. A candidate solution, denoted $s(z)$ (e.g., a query plan), is represented by an encoding $z\in\{0,1\}^{n}$ in the quantum solution space, where $n$ is the number of binary decision variables in the encoding. The objective is to maximize solution quality while adhering to a strict budget for quantum processing time.

\noindent\textbf{{Performance.}}
We measure solution quality by the performance of executing query $\mathcal{Q}$ on database $\DB$. Specifically,
\[
\max \,\Performance(s;\mathcal{Q};\DB)\ \Longleftrightarrow\ \min \,\Latency(s;\mathcal{Q};\DB),
\]
\emph{where} $\Latency(s;\mathcal{Q};\DB)$ is the execution time of $\mathcal{Q}$ under solution $s$ on $\DB$, so a lower latency indicates a higher quality.

\begin{figure*}[t]
  \centering
  \begin{subfigure}[t]{0.50\linewidth}
    \centering
    \includegraphics[width=\textwidth]{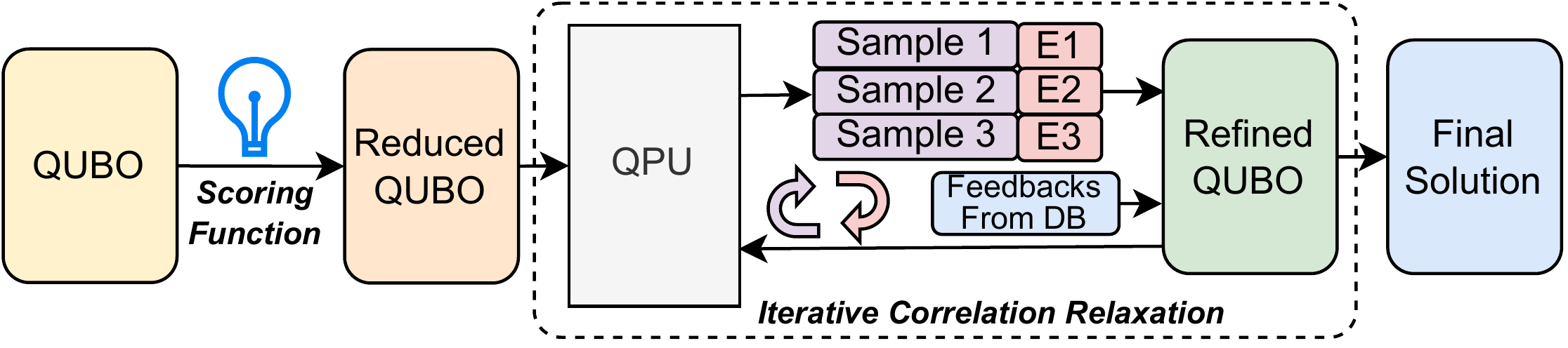}  
    \caption{Iterative Correlation Relaxation.}
     \label{fig:iter-corr-relax}
  \end{subfigure}
  \hfill
  \begin{subfigure}[t]{0.48\linewidth}
    \centering
    \includegraphics[width=\textwidth]{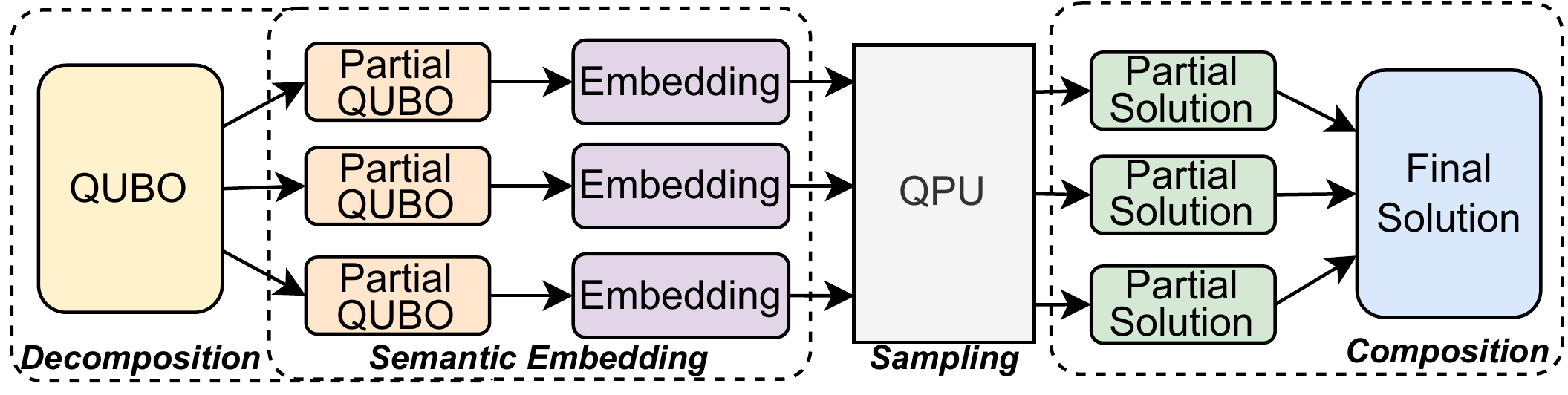} 
    \caption{Problem-Aware Decomposition and Composition.}
     \label{fig:problem-aware-decomposition}
  \end{subfigure}
  \caption{Scalable Hybrid Quantum-Classical Methods.}
\end{figure*}

\noindent\textbf{{Quantum Lifecycle.}}
This quantity represents the time consumed by our quantum computing pipeline to obtain the candidate solution $z$. To reach a high-quality solution, the algorithm may interact with the quantum hardware multiple times (see Section~\ref{subsec:research_task_1}). Let $N$ be the number of iterations. In iteration $i$, let $T_{Q_i}$ denote the quantum solver compute time, $T_{C_i}$ the communication time between the DBMS and the quantum service, and $T_{R_i}$ the client-side QUBO refinement time. The total quantum lifecycle time is
\[
\QunatumTime(z)\;=\;\sum_{i=1}^{N}\bigl(T_{Q_i}+T_{C_i}+T_{R_i}\bigr),
\]
where $\QunatumTime(z)$ is the end-to-end time spent within the quantum lifecycle to produce $z$.

\noindent\textbf{Performance Maximization under a Time Budget.}
Given a time budget $\tau>0$, our goal is to find, within this budget, the solution that yields the lowest query execution latency. The problem is formulated as
\[
\begin{aligned}
\textbf{Given: } & \mathcal{Q},\ \DB,\ \tau\in\mathbb{R}_{>0} \\
\textbf{Decisions: } & z\in\{0,1\}^{n},\ s=s(z) \\
\textbf{Minimize: } & \Latency(s;\mathcal{Q};\DB) \\
\textbf{s.t. } & \QunatumTime(z)\ \le\ \tau \qquad \ \ (\textit{Time budget})\\
& g_j(s)\ \le\ 0,\ \ j=1,\ldots,m\ \ (\textit{Original constraints})
\end{aligned}
\]

\section{Scalable Hybrid Quantum-Classical Methods for Large-Scale Database Optimization}
\label{sec:scalable_hybrid_quantum_classical}

In this section, we present two approaches. Both address the first challenge (C1) by avoiding reliance on black box solvers and by providing transparent control over the quantum hardware. The first approach addresses the second challenge (C2) of QUBO overcomplexity, enabling a tunable trade-off between efficiency and accuracy. It introduces an iterative correlation relaxation framework that retains the full set of variables while pruning and progressively reintroducing correlations to simplify QUBOs (Section~\ref{subsec:research_task_1}). The second approach addresses the third challenge (C3), where the QUBO exceeds current hardware capacity. It adopts a problem-aware decomposition and composition strategy that preserves both variables and correlations, partitions the QUBO into smaller, semantically meaningful subproblems, solves them independently, and then reconciles the results (Section~\ref{subsec:research_task_2}).

\subsection{Iterative Correlation Relaxation}
\label{subsec:research_task_1}
Selectively relaxing correlations between variables yields a sparser interaction structure, which simplifies mapping correlation onto the quantum annealer for overly complex QUBO instances and naturally enables more efficient execution. With proper correlation relaxation, even though the reduced QUBO omits several correlations, it can still produce solutions of comparable quality. In this case, we propose an iterative correlation relaxation framework to address these limitations by balancing QUBO sparsity (i.e., over-relaxation leads to a sparse QUBO) and solution accuracy (i.e., limited relaxation maximally maintains the original QUBO features) through problem-specific heuristics and feedback-driven refinement. This iterative mechanism also provides flexibility in controlling the trade-off between the solution accuracy and latency from quantum computing. 

Fig.~\ref{fig:iter-corr-relax} shows a complete framework. We first obtain a reduced QUBO that preserves only the correlations with the largest estimated influence from the initial QUBO. These correlations are selected using \texttt{scoring functions} tailored to each database problem. For example, in join order optimization, we prioritize correlations between joins that yield large intermediate result cardinalities~\cite{joheuristic23}. For transaction scheduling, we retain edges that involve transactions with frequent contention~\cite{transconflict24}. For index tuning, we emphasize access pattern correlations among indexes that are often co-accessed across queries~\cite{indextunerl22}. In these cases, the scoring functions yield higher scores for these correlations, which serve crucial roles in the original database optimization problem.

The QPU, e.g., the D-Wave quantum annealer~\cite{dwaveleap}, is used as a parameterized sampler~\cite{qbg18} that produces candidate solutions (i.e., samples and corresponding energies) under the reduced correlation set with the current coefficients \(h\) and \(J\) (introduced in Section~\ref{subsec:quantum-annealing}). The empirical distribution of these samples provides the first source of feedback. Additionally, closely integrated with the actual database system, this provides more feedback, such as solution quality metrics (e.g., current query plan cost in the join order problem) and constraint violation statistics (e.g., the number of invalid join subplans). These signals are analyzed to identify missing correlations that have a significant impact on actual query execution performance. The identified correlations, i.e., the pairwise terms in $J$, are selectively reintroduced in subsequent iterations, thereby progressively refining the QUBO structure.

The iterative procedure terminates when any of the following conditions is met: (i) solution quality stabilizes, e.g., no further cost reduction over several consecutive iterations; (ii) the energy landscape reaches a predefined stability threshold; or (iii) the maximum number of iterations is reached.

\noindent{\textit{\textbf{Convergence Strategies.}}}
In iterative QUBO refinement, a central challenge is to devise an effective strategy that ensures convergence. Simply reintroducing correlations in a greedy manner may lead to oscillations or premature convergence, especially when the database optimization problem exhibits a dense cost structure. To address this challenge, we minimize the Kullback-Leibler (KL) divergence~\cite{kldiv51} between the ideal Boltzmann distribution~\cite{qbg18,Bian2016MappingCO} over valid solutions and the empirical distribution generated by the QPU. Aligning these distributions approximates the energy landscape of the fully constrained problem. The resulting iterative updates resemble machine learning based gradient descent, progressively reducing the discrepancy between the two distributions. As complementary options, entropy-based regularization~\cite{jsd91} can serve as fallbacks when KL-divergence minimization exhibits instability on specific workloads.

\subsection{Problem-Aware Decomposition and Composition}
\label{subsec:research_task_2}
To address the third challenge, where the size of a QUBO exceeds quantum hardware capacity, we preserve the full problem structure by retaining all variables and pairwise interactions, while decomposing the problem into smaller subproblems that are solved individually and then recomposed (e.g.,~\cite{dwaveleap,fujitsu2023,qiskit23,qtranssched25,subquboonehot25,clustersubqubo25,partitionencode17}). We propose a problem-aware hybrid optimization pipeline that fully considers the key semantics of the underlying database optimization problem. We focus on join order optimization~\cite{cqmjo24,sabekbushyjoin25,DA} and transaction scheduling~\cite{transsched24,transactionqubo} as representative use cases. The pipeline is generalizable to other problems. It comprises four stages, detailed as follows.

\subsubsection{\textbf{Decomposition Stage}}
In this stage, large QUBOs are divided into smaller subproblems to reduce inter-partition dependencies while preserving all correlations relevant to database operation semantics. For join order optimization, we leverage the join graph representation introduced in~\cite{joinsearchspace01}, which encodes candidate binary join operations as nodes (i.e., QUBO variables) and uses edges (i.e., QUBO correlations) to indicate the presence of shared base relations. Each spanning tree of this graph corresponds to a valid join plan, and the weights on edges and vertices capture join costs and selectivity factors. Based on this model, we design decomposition heuristics that minimize the number of shared join keys across partitions, thereby enabling partial plans to be solved independently while retaining cost accuracy.

For transaction scheduling, we partition according to contention profiles~\cite{chiller20}, grouping transactions that frequently contend and isolating independent ones to enable parallel solving while preserving correctness. A central challenge in both settings is selecting the right decomposition granularity. Partitions that are too coarse may exceed hardware capacity, whereas partitions that are too fine increase the complexity of future composition. We address this by developing adaptive decomposition strategies guided by workload statistics, such as join selectivity and transaction contention levels, to balance subproblem size and future composition cost.

\subsubsection{\textbf{Embedding Stage}}
In this stage, we map the variables of each decomposed subproblem onto the QPU while preserving the \texttt{semantic locality} of the underlying database operation. This is unlike generic minor embedding methods~\cite{dwaveadvantage}, which often produce long and fragile chains. Chains group multiple physical qubits into a superqubit to provide stronger effective connectivity than couplings between single qubits. However, chains are fragile in practical quantum execution, which makes them ill-suited and unreliable for representing the logical relationships inherent in the problem. For join order optimization, we embed all binary join operations that involve a given base table to reflect their shared data paths. For transaction scheduling, we map critical conflict pairs to low-noise regions of the device to reduce chain breaks. We maintain a library of reusable embedding templates (e.g., for cyclic join patterns~\cite{DA, readytoleap}), which can be adapted to different hardware architectures, such as D-Wave Pegasus~\cite{Pegasus} and Zephyr~\cite{zephyr}.

\subsubsection{\textbf{Sampling Stage}}
In this stage, we apply QPU sampling strategically, adapting to the characteristics of each decomposed subproblem in order to obtain high-quality solutions. For example, for subproblems in join ordering that involve many relations or large intermediate results, we allocate longer annealing time or adopt adaptive annealing schedules to allow the QPU to explore the solution space more thoroughly. For low complexity or less critical subproblems, we use shorter runs to improve efficiency. Current quantum hardware often uses static parameters for all QPU tasks, which leads to inefficiency when the subproblem complexity varies~\cite{dwaveleap,fujitsu2023}. This strategy employs adaptive sampling policies that tune annealing time, the number of reads, and bias schedules based on subproblem cost estimates and solver feedback (e.g., energy variance). By allocating resources according to subproblem difficulty and updating parameters from feedback, the sampling strategy makes better use of existing quantum hardware and improves performance while adaptively handling large QUBOs.

\subsubsection{\textbf{Composition Stage}} 
In this stage, we employ a two-phase coordination framework to reconcile partial solutions into a globally consistent result. In the first phase, we apply domain-aware belief propagation, inspired by prior work on structured Ising models and domain decomposition~\cite{qdiscrete14,Bian2016MappingCO,qbg18}. These techniques pass messages over overlapping variables to reconcile local optima and steer convergence toward a coherent global solution. For join order optimization, for instance, partitions that share join keys exchange consensus updates to align cost estimates. In the second phase, we apply post-processing heuristics such as very large neighborhood search~\cite{lNS19} and tabu search~\cite{partitionencode17} to iteratively refine the merged solution, particularly for transaction scheduling workloads where reconciliation must respect global constraints, including lock ordering and latency budgets. These methods serve as fallback procedures when belief propagation is ineffective or computationally expensive, and they enable incremental improvements without requiring a full re-optimization.

\begin{figure*}[t]
    \centering
    \includegraphics[width=0.9\linewidth]{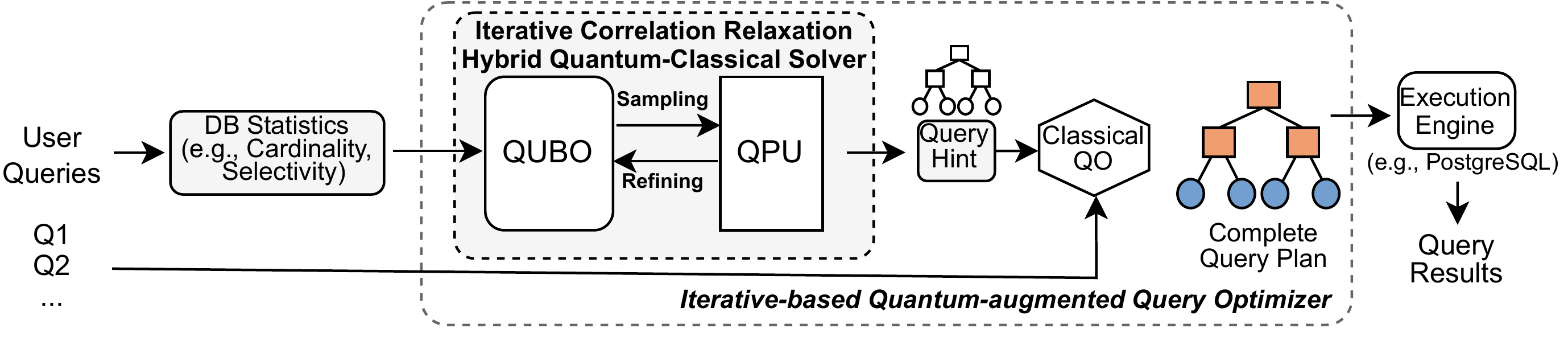}
    \caption{Framework of Iterative-based Quantum-augmented Query Optimizer.}
    \label{fig:q2o}
\end{figure*}

\section{Preliminary Prototype: Quantum-augmented Query Optimizer}\label{sec:prototype}
In our preliminary prototype, we integrate the iterative correlation relaxation approach (Section~\ref{subsec:research_task_1}) with the PostgreSQL query optimizer~\cite{postgres}. Fig.~\ref{fig:q2o} illustrates the framework of the iterative quantum augmented query optimizer, referred to as {\oursystem}, which serves as the initial prototype for tightly coupling our proposed approaches with the database system.

Upon receiving a user query, {\oursystem} first retrieves cardinality and selectivity statistics from the DBMS engine. These statistics are used to construct the QUBO formulation, which parameterizes both the objective (e.g., cost function) and the constraints (e.g., join tree structure). The expected output, the join order sequence, is encoded in the decision variables within the QUBO. Once the QUBO is constructed, denoted \texttt{Initial} QUBO, it is submitted to the QPU (i.e., a quantum annealer in our initial prototype). The QPU performs quantum sampling and returns $n$ candidate solutions, along with their corresponding energies. The distribution of these $n$ samples and their energies is then used to refine \texttt{Initial} QUBO into \texttt{Refined} QUBO according to the methods proposed in Section~\ref{subsec:research_task_1}. The combined sampling and refinement step is treated as one iteration. The available time budget determines the number of iterations, which may be specified directly by the user or determined automatically by the system. After the prescribed iterations, the iterative-based hybrid quantum-classical solver returns a join order sequence that is the best found under the timing constraint.

This sequence is then translated into a query plan hint, which is used to guide a classical query optimizer in producing a complete query plan\footnote{This method is commonly used in many well-known query optimizers (e.g., PostgreSQL~\cite{postgresql}, MySQL~\cite{mysql}, Oracle~\cite{oracle}) and learned query optimizers (LQOs) (e.g.,~\cite{liu2025serag,liu2025sefrqo}).}. In our preliminary prototype, we use \textit{pg\_hint\_plan}~\cite{pg_hint_plan} to tweak PostgreSQL execution plans by embedding hints in SQL comments. The join order is specified with a clause beginning with a key word \texttt{Leading}; inner parentheses denote grouping precedence in the join order sequence. For example, the bushy join hint \texttt{Leading((a(b c))(d e))} indicates that $b \bowtie c$ is performed first, followed by $a \bowtie (b \bowtie c)$, then $d \bowtie e$, and finally the two intermediate results are joined. PostgreSQL then selects the join operators (e.g., hash join, nested loop join) for each join. Once the optimizer accepts the hint, it is converted into a complete query plan, which the execution engine runs to produce and return the query results.

\section{Preliminary Experiments}\label{sec:results}
In this section, we evaluate our preliminary prototype (introduced in Section~\ref{sec:prototype}), which applies 
the iterative correlation relaxation approach (introduced in Section~\ref{subsec:research_task_1}) to solve the query optimization problem in a real-time database system.
We first present our experimental setup in Section~\ref{subsec:setups}. We then evaluate our approach by addressing the following questions: (1) How effective is {\oursystem} for query optimization, as measured by query execution latency, compared with a classical query optimizer? (2) How does {\oursystem} compare with a black-box quantum solver in terms of efficiency and solution quality? (3) What is the end-to-end latency of the approach, and does it maintain an advantage when communicating with an external quantum service? (4) What is the detailed timing breakdown on the quantum hardware during the core quantum computing process?

\begin{figure*}[t]
  \centering
  \begin{subfigure}[t]{0.63\linewidth}
    \centering
    \includegraphics[width=\linewidth]{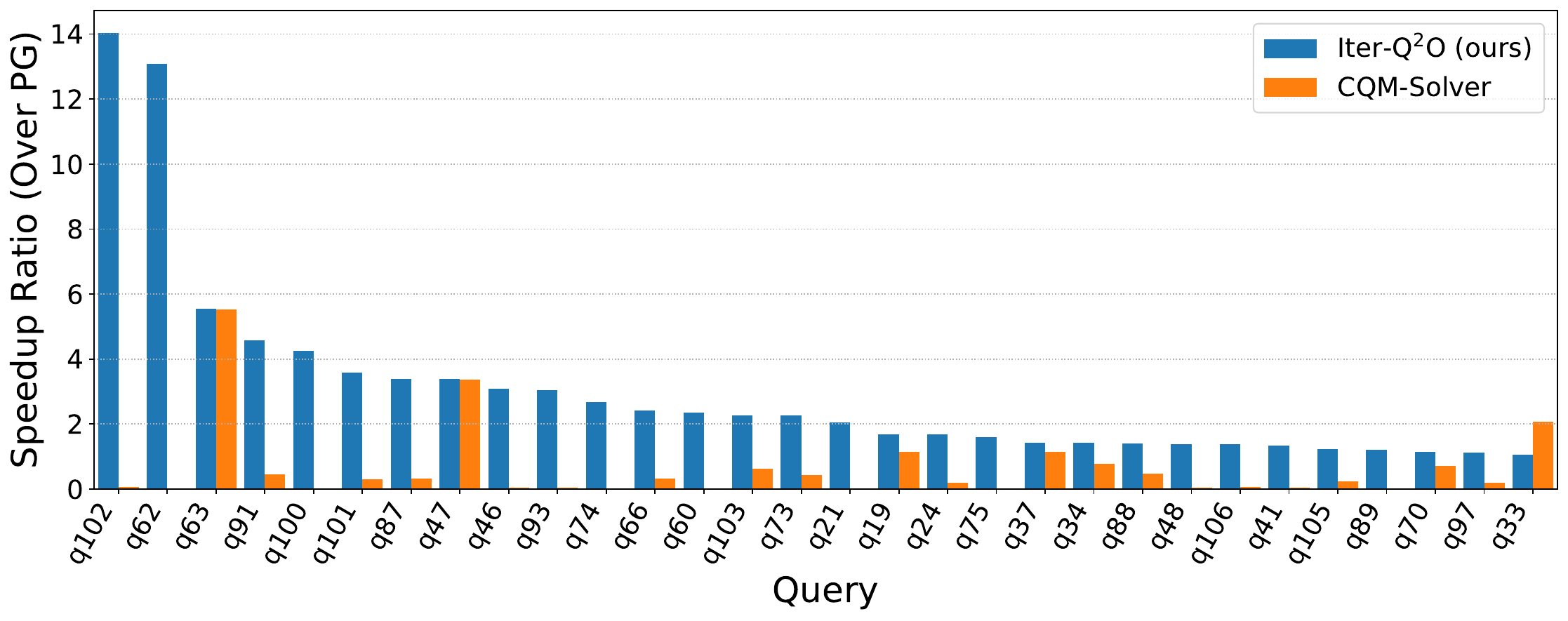}
    \caption{{\oursystem} and CQM-Solver (on JOB)}
    \label{fig:job_speedup}\label{fig:compare_cqm}
  \end{subfigure}\hfill
  \begin{subfigure}[t]{0.35\linewidth}
    \centering
    \includegraphics[width=\linewidth]{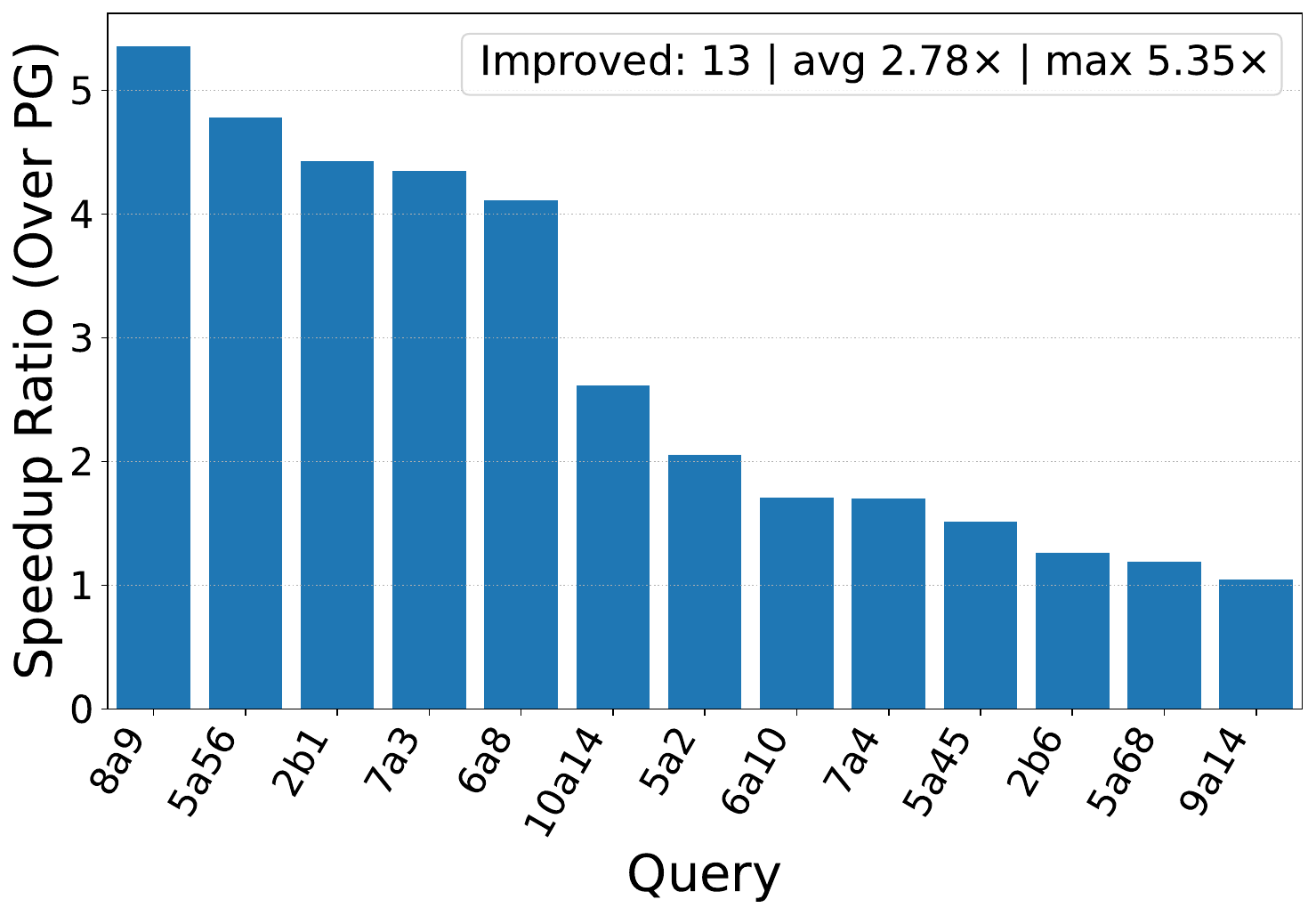}
    \caption{{\oursystem} (on CEB)}
    \label{fig:ceb_speedup}
  \end{subfigure}
  \caption{A Selection of Query Plans Generated by Our Approach Outperforms PostgreSQL's Default Plans (Iteration = 2).}
  \label{fig:job_ceb_speedup}
\end{figure*}

\subsection{Experimental Setup}\label{subsec:setups}
We run all our experiments on a single server equipped with an Intel Core i9-9980XE CPU and 128 GB of RAM. We employ the Advantage2\_system1.7 as the quantum sampler on the D-Wave Leap quantum service platform. We set up PostgreSQL {$16.4$} with the corresponding \textit{pg\_hint\_plan} version. We utilize D-Wave's CQM-Solver~\cite{dwavehss}, a representative black-box quantum solver, as a baseline for black box approaches. We evaluate our approach using the \textit{Join Order Benchmark (JOB)}~\cite{qoeval15} and the \textit{Cardinality Estimation Benchmark (CEB)}~\cite{flowloss}. Both JOB and CEB are constructed on the IMDB movie database schema and data, with CEB containing more complex query patterns and a wider range of cardinality estimation challenges than JOB. This complexity makes JOB and CEB suitable and widely adopted benchmarks for testing join order optimization problems. 

In the following experiment, when testing the preliminary iterative-based quantum-augmented query optimizer introduced in Section~\ref{sec:prototype}, we set the number of QUBO sampling-and-refinement iterations to two to invoke the iterative correlation relaxation. In this case, the complete lifecycle (in our 2-iteration setup) comprises: (i) generating the problem from query-related statistics; (ii) connecting to the quantum solver; (iii) submitting the problem and retrieving results; (iv) performing local QUBO refinement; and (v) resubmitting the problem and retrieving the final results.

\subsection{Experimental Results}\label{subsec:pre-results}

\noindent\textbf{Query Plan Quality.}
We evaluate the query plan quality using the actual execution time of JOB and CEB queries on PostgreSQL. Fig.~\ref{fig:job_ceb_speedup} compares the latencies of plans generated by our method with those produced by PostgreSQL. On the JOB workload (113 queries), our approach achieves speedups on 36 queries, with a maximum speedup of 14.02x and an average speedup of 2.70x. CEB queries are used to investigate {\oursystem}'s capability of handling complex workload. On the CEB workload (40 queries), our approach achieves speedups on 13 queries, with a maximum speedup of 5.35x and an average speedup of 2.78x. For the remaining queries, our approach showed only limited performance degradation.

\begin{table}[t]
\caption{End-To-End Latency Breakdown and Comparison Between Our Approach and PostgreSQL.}
\renewcommand{\arraystretch}{1.2}
\resizebox{\linewidth}{!}{
\begin{tabular}{c|c|cc|c}
\hline
 & \textbf{Workload} & \multicolumn{2}{c|}{JOB} & CEB \\ \hline
 & \textbf{Time (ms)} & q102 & q62 & 8a9 \\ \hline
\multirow{2}{*}{\textbf{PG}} & Planning & 3172.32 & 17.33 & 82.33 \\ \cline{2-5} 
 & Execution & 8039.26 & 5177.80 & 5992.39 \\ \hline
\multirow{3}{*}{\textbf{CQM}} & Planning with hints & 2467.99 & 4.54 & 43.89 \\ \cline{2-5} 
 & Execution & 44968.41 & 311489.41 & 512241.75 \\ \cline{2-5} 
 & Quantum Lifecycle & 9092.41 & 7591.69 & 9002.50 \\ \hline
\multirow{3}{*}{\textbf{\begin{tabular}[c]{@{}c@{}}$\oursystem$\\ (ours)\end{tabular}}} & Planning with hints & 2548.51 & 7.31 & 65.91 \\ \cline{2-5} 
 & Execution & \textbf{573.27} & \textbf{395.74} & \textbf{1119.09} \\ \cline{2-5} 
 & Quantum Lifecycle & 3168.31 & 1553.35 & 1772.95 \\ \hline
\multirow{3}{*}{\textbf{\begin{tabular}[c]{@{}c@{}}Our Gain\\ (Over PG)\end{tabular}}} & \textbf{Absolute Time Saving} & \textbf{4921.49} & \textbf{3238.73} & \textbf{2957.95} \\ \cline{2-5} 
 & \textbf{Exec Time $\uparrow$} & \textbf{14.02x} & \textbf{13.08x} & \textbf{5.35x} \\ \cline{2-5} 
 & \textbf{E2E Latency $\uparrow$} & \textbf{1.78x} & \textbf{2.65x} & \textbf{2.05x} \\ \hline
\end{tabular}}\label{tab:latency}
\end{table}

\noindent\textbf{Comparison with Black-box Quantum Solver.}
For a fair comparison, we apply the same initial QUBO formulation to both our prototype and D-Wave's CQM-Solver. Fig.~\ref{fig:compare_cqm} reports the speedup over PostgreSQL for both methods. For most queries, our approach outperforms the black-box solver. The better performance arises because our iterative solver focuses on the correlations that matter and iteratively corrects the relaxation by reintroducing important correlations. In contrast, we cannot provide problem-aware, fine-grained controls with the CQM-Solver, as it operates as a black box.

Moreover, Table~\ref{tab:latency} illustrates that the CQM-Solver incurs a much larger overhead during the quantum lifecycle. This is due to the CQM-Solver enforcing a 5-second minimum time limit before returning results. In contrast, our approach relies solely on the lowest-level operation, quantum sampling, which enables us to circumvent the constraints imposed by black-box quantum solvers.

\noindent\textbf{End-to-End (E2E) Latency.}
We evaluate End-to-End latency to compare the complete query execution pipeline. For PostgreSQL, E2E latency consists of planning time and execution time. For our approach, E2E latency includes PostgreSQL planning time (with query hint generated by our approach), PostgreSQL execution time, and the time consumed by the complete quantum lifecycle. Table~\ref{tab:latency} reports a detailed breakdown of these components for two JOB queries and one CEB query. Although our approach incurs additional quantum overhead, primarily due to cloud service communication, it generates better join order sequences than PostgreSQL's defaults, resulting in query plans that reduce execution time and offset this overhead. All reported results are obtained on a local IMDB dataset. As data size increases, the absolute execution time improvement is expected to scale, while the quantum overhead will not scale correspondingly. Overall, our method achieves E2E latency reductions on the order of seconds (i.e., 2957.95\,ms, 3238.73\,ms and 4921.49\,ms), yielding 1.78x to 2.65x E2E latency speedups.

\noindent\textbf{Detailed Timing Breakdown of the Quantum Lifecycle.}
In Table~\ref{tab:latency}, the JOB query q102 spent 3168.31\,ms in the quantum lifecycle. We further quantify the portion directly correlated to the quantum hardware itself. 

\begin{table}[t]
  \centering
  \caption{Timing Breakdown on Quantum Service (ms).}
  \label{tab:timing_breakdown}
  \resizebox{\linewidth}{!}{%
  \begin{tabular}{lrrrr}
    \toprule
    \textbf{Metric} & \textbf{Iter1} & \textbf{Iter2} & \textbf{Total} & \textbf{Avg} \\
    \midrule
    Ingress (created$\rightarrow$received) & 87.896 & 77.505 & 165.401 & 82.701 \\
    Solve (received$\rightarrow$solved)    & 361.214 & 246.652 & 607.866 & 303.933 \\
    Egress (solved$\rightarrow$resolved)   & 122.164 & 122.176 & 244.340 & 122.170 \\
    \cmidrule(lr){1-5}
    \textbf{End-to-end (created$\rightarrow$resolved)} & \textbf{571.274} & \textbf{446.333} & \textbf{1017.607} & \textbf{508.803} \\
    \midrule
    QPU programming                        & 15.760 & 15.760 & 31.520 & 15.760 \\
    QPU sampling                           & 26.356 & 26.356 & 52.712 & 26.356 \\
    QPU access time (reported)             & 42.116 & 42.116 & 84.232 & 42.116 \\
    QPU access overhead                    & 0.822 & 0.798 & 1.620 & 0.810 \\
    Post-process overhead      & 0.037 & 0.036 & 0.073 & 0.036 \\
    \bottomrule
  \end{tabular}%
  }
\end{table}

Table~\ref{tab:timing_breakdown} provides a detailed timing breakdown for two iterations of interaction with the quantum annealer. Across the two iterations, the end-to-end times are 571.274\,ms and 446.333\,ms (average 508.803\,ms). This end-to-end time includes all time consumed on the commercial quantum cloud platform from when the problem is created (i.e., \texttt{created}) until the solution is returned from the quantum hardware to the quantum cloud platform (i.e., \texttt{resolved}). The end-to-end quantum solving time of about 500\,ms is encouraging. We also observe that the total time of these two iterations consumed on the quantum cloud platform is 1017.607\,ms, which accounts for only a small portion of the client-side observed 3168.31\,ms (32.12\%). The remaining time is mostly due to multiple bi-directional cloud communications between our database platform and the commercial cloud platform.

On the quantum hardware itself, the average QPU programming and sampling times are 15.760\,ms and 26.356\,ms, respectively. The reported QPU access time is 42.116\,ms, which represents the time consumed on the core quantum computing process. The QPU access overhead and post-processing overhead are relatively small. Overall, these measurements indicate the current status of solving time and overhead on the quantum hardware. Although most time is spent on cloud communication, the observed end-to-end latency of the quantum hardware and the complete quantum lifecycle is practical for solving real-world and even real-time database problems.

\noindent\textbf{Challenges and Future Improvement.}
In each iteration of our proposed iterative framework, we refine the QUBO locally based on the obtained samples and energies, and then submit the refined QUBO. Each retrieval and submission incurs non-trivial cloud communication, as shown in Section~\ref{subsec:pre-results}, with around 60\% of the quantum lifecycle time spent on communication. To reduce this overhead, we propose moving the refinement loop to the server side, for example, by compiling our proposed framework and deploying it as close as possible to the quantum hardware, which would substantially improve runtime efficiency and make the approach better suited to real-world and real-time problem solving.

\section{Conclusion} \label{sec:conclusion}
In this paper, we propose the first real-time quantum-augmented database system that tightly integrates a hybrid quantum-classical optimizer. This system enables transparent solutions rather than solely relying on black-box quantum solvers. We then present two scalability strategies targeting large-scale challenges. To address the QUBO overcomplexity challenge, we introduce an iterative correlation relaxation framework that simplifies the QUBO structure while balancing accuracy and efficiency. To address the QUBO oversizing challenge, we propose a problem-aware decomposition and composition pipeline that leverages database semantics to partition QUBOs into subproblems, thereby enabling the solution of large-scale optimization problems. Our preliminary prototype and evaluations demonstrate improved performance over both classical database components and a black-box quantum solver. Moving forward, we aim to design more efficient and accurate quantum augmented systems for large-scale database optimization that are practical in real-time settings.

\sloppy
\bibliographystyle{IEEEtran}
\bibliography{refs/ml4db,refs/qc4db,refs/references,refs/spatialmln}

\end{document}